%% file: main.tex
\documentclass[reprint,superscriptaddress,amsmath,amssymb, aps,prl]{revtex4-2}
\usepackage{standalone}
\usepackage{import}
\usepackage{graphicx}
\usepackage{dcolumn}
\usepackage{bm}
\usepackage{url}
\usepackage{hyperref}
\usepackage{xcolor}
\usepackage{commath}
\usepackage{braket}
\usepackage{dsfont}
\usepackage[T1]{fontenc}
\usepackage{siunitx}
\usepackage{verbatim}
\usepackage{subfiles}

\newcommand{\hint}{H_\text{int}}
\newcommand{\imi}{\mathrm{i}}
\newcommand{\adagger}{a^\dagger}
\newcommand{\dt}{\Delta t}
\newcommand{\sigmap}{\sigma_+}
\newcommand{\sigmam}{\sigma_-}
\newcommand{\ketbra}[2]{\ensuremath{\ket{#1}\hspace{-1pt}\bra{#2}}}
\newcommand{\expup}[1]{\mathrm{e}^{#1}}

\newcommand{\nmeas}{N_\text{bit}}

\newcommand{\psin}[1]{\ket{\psi_{#1}}}
\newcommand{\filter}[1]{\mathcal{F}_{#1}}
\newcommand{\ntrajs}{N_\text{trajs}}
\newcommand{\figpanel}[2]{\hyperref[#1]{\ref*{#1}(#2)}}
\renewcommand{\angle}{\phi_\text{int}}
\newcommand{\swapangle}{\phi_\text{SWAP}}

\newcommand{\phase}{\theta}

\makeatletter
\AtBeginDocument{\let\LS@rot\@undefined}
\makeatother

\makeatletter
\def\maketitle{
\@author@finish
\title@column\titleblock@produce
\suppressfloats[t]}
\makeatother

\begin{document}

\title{Digital homodyne and heterodyne detection for stationary bosonic modes}

\author{Ingrid Strandberg}
\affiliation{Department of Microtechnology and Nanoscience MC2, Chalmers University of Technology, SE-412 96
G\"oteborg, Sweden}

\author{Axel Eriksson}
\affiliation{Department of Microtechnology and Nanoscience MC2, Chalmers University of Technology, SE-412 96
G\"oteborg, Sweden}

\author{Baptiste Royer}
\affiliation{Département de Physique and Institut Quantique, Université de Sherbrooke, Sherbrooke J1K 2R1 QC, Canada}

\author{Mikael Kervinen}
\email{Present address: VTT Technical Research Centre of Finland Ltd. Tietotie 3, Espoo 02150, Finland}
\affiliation{Department of Microtechnology and Nanoscience MC2, Chalmers University of Technology, SE-412 96
G\"oteborg, Sweden}

\author{Simone Gasparinetti}
\affiliation{Department of Microtechnology and Nanoscience MC2, Chalmers University of Technology, SE-412 96
G\"oteborg, Sweden}

\begin{abstract}

Homo- and heterodyne detection are fundamental techniques for measuring propagating electromagnetic fields. However, applying these techniques to stationary fields confined in cavities poses a challenge.
As a way to overcome this challenge, we propose to use repeated indirect measurements of a two-level system interacting with the cavity. We demonstrate numerically that the proposed measurement scheme faithfully reproduces measurement statistics of homo- or heterodyne detection at the single-shot level. The scheme can be implemented in various physical architectures, including circuit quantum electrodynamics. Our results pave the way to the implementation of quantum algorithms requiring linear detection, including quantum verification protocols, in stationary modes.
\end{abstract}

\maketitle

\section{Introduction}

Continuous monitoring of quantum states of light has a long history in quantum optics. Direct photon detection~\cite{Short1983Aug,Rarity1987May} and homodyne detection~\cite{Slusher1985Nov, Wu1986Nov} were used to reveal nonclassical properties of the electromagnetic field already in the 1980s. These two methods, along with heterodyne detection, are the basic techniques for detecting optical radiation~\cite{Leonhardt1997-kr, Bachor2004-eq}. Since the 2000s, with the advent of circuit quantum electrodynamics~\cite{Blais2004Jun} as a platform for quantum information processing, there has been an increased interest in detecting microwave radiation at the quantum level. While photon number resolving detection of propagating microwave photons is challenging due to their low energy~\cite{Dassonneville2020Oct}, homodyne and heterodyne detection of propagating fields can be performed thanks to the availability of low-noise linear amplifiers~\cite{Eichler2012Sep}. The ability to perform these measurements is of interest since they suffice to implement any multimode Gaussian operation in continuous-variable measurement-based quantum computation~\cite{Menicucci2006Sep, Gu2009Jun}, boson sampling~\cite{Chakhmakhchyan2017Sep} as well as efficient verification of it~\cite{Chabaud2021Nov}, and reliable state verification of an untrusted preparation~\cite{Aolita2015Nov, chabaud_et_al:LIPIcs:2020:12062}. On the other hand, these types of measurements are not straightforward to perform on confined cavity fields, which are important in the context of quantum computing with bosonic modes~\cite{Sivak2023Apr,Ofek2016Aug,Joshi2021Apr}. 
Information about a cavity field can be obtained by measuring the field leaking out of the cavity, but since photon loss is a major obstacle for bosonic quantum information processing, cavities with as low as possible loss rate is generally desired, which makes monitoring the leaked output inefficient. One way to overcome this problem is to swap the stationary mode with a propagating mode~\cite{Pfaff2017Sep}. However, this solution requires tunable couplers which are difficult to engineer and require added hardware. An alternative is to indirectly probe the cavity field. This type of indirect measurement was first performed to measure the photon number inside a cavity by letting Rydberg atoms cross it and measuring the atoms afterward~\cite{Brune1990Aug}. Similar ideas have subsequently been used in superconducting circuits, where a qubit has not only been used as a probe for measurement of the photon number~\cite{Schuster2007Feb}, but also the cavity Wigner function~\cite{Banaszek1999Jul, Nogues2000Oct, Bertet2002Oct}. However, to-date, highly efficient homo- or heterodyne detection measurements of cavity modes are lacking.

In this Letter, we propose using a sequence of indirect measurements, assisted by an ancillary qubit, to perform a digital version of homodyne and heterodyne detection of a stationary bosonic mode.
For this reason, we refer to our measurement protocol as \emph{qubitdyne} detection.
We demonstrate by numerical calculations that qubitdyne reproduces the measurement statistics associated with the fundamental types of quadrature measurements in quantum optics: homodyne- and heterodyne detection.
The simple, bilinear interaction Hamiltonian needed to perform the qubitdyne protocol can be implemented in a variety of systems: trapped ions and atoms~\cite{Borne2020Apr, Bechler2018Oct}, nanomechanical oscillators~\cite{Wollack2022Apr}, NV centers~\cite{Liu2016Nov}, and superconducting circuits~\cite{Blais2004Jun}.

\section{Repeated indirect measurements setup}

Our qubitdyne setup can be described as a realization of a so-called \emph{repeated quantum interactions model}~\cite{Attal2007Mar, Bruneau2006Oct} or \emph{collision model}~\cite{Ciccarello2017Dec, Cusumano2022Sep, Campbell2021May}.
In the general model, a quantum system is in contact with an environment represented as a chain of independent smaller systems called probes. The time evolution is obtained by consecutive interactions with each probe during a short time interval $\dt$, and after each interaction a measurement is performed on the corresponding probe~\cite{Pellegrini2010Nov,Ashida2020Jan}, which provides indirect information of the cavity state.

\begin{figure}[h!]
    \centering
    \includegraphics[width=\columnwidth]{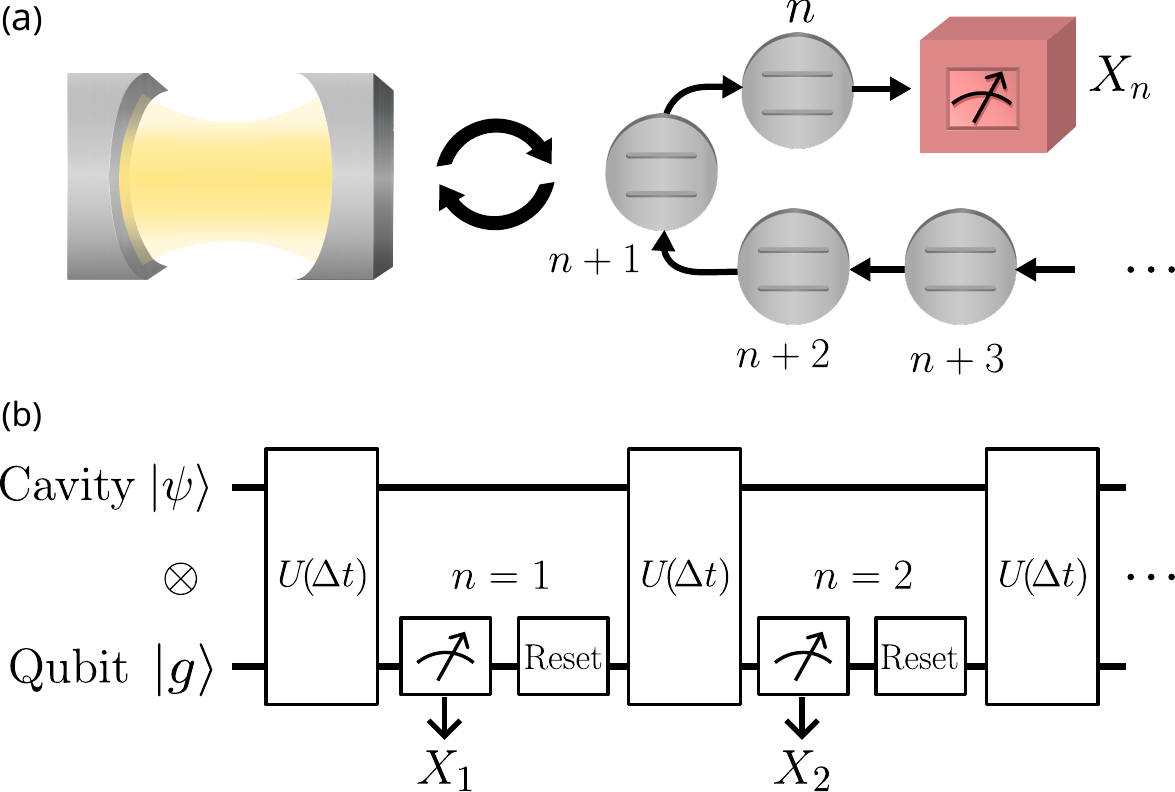}
    \caption{The system realizing digital homo- and heterodyne detection of a stationary mode. (a) Interaction and subsequent measurement of independent qubits interacting with a cavity, with measurement outcome $X_n=\pm 1$ for the $n$th qubit. (b) Circuit diagram representation of the procedure. The total system evolves with a unitary operator $U(\dt)$ during a time interval $\dt$, corresponding to a partial SWAP. A projective measurement is performed on the qubit, after which it is reset to its ground state. The process is repeated and a sequence of measurement outcomes $\{X_1, X_2, \ldots,X_{\nmeas}\}$ is obtained.}
    \label{fig:system}
\end{figure}

In our case, the primary system is a long-lived cavity. If the interaction time is sufficiently short and the coupling weak, then the probability of transferring more than one photon from the cavity to a probe is negligible, meaning that each probe can be represented as a two-level system, i.e., a qubit~\cite{Brun2002Jun}. Such a repeated interactions model is illustrated in Fig.~\figpanel{fig:system}{a}. This type of model reproduces open quantum system dynamics~\cite{Attal2007Mar, Bruneau2006Oct,Ciccarello2017Dec, Cusumano2022Sep, Campbell2021May, Ciccarello2022Apr}, and has been used for investigating systems with complex environments~\cite{Dabrowska2019Feb, Daryanoosh2022Aug, Ciccarello2013Apr}.

To simplify the realization of the model, instead of a chain of qubits, we consider a single qubit that is reset to its ground state after each measurement.
The quantum circuit realizing our scheme is drawn in Fig.~\figpanel{fig:system}{b}.
The cavity and qubit modes are represented by creation and annihilation operators $\adagger$, $a$ and $\sigmap$, $\sigmam$, respectively.
With a Jaynes-Cummings coupling between the systems, the unitary evolution for an interaction of time $\dt$ is
\begin{equation}\label{eq:unitary}
    U = \exp\left(-\imi \angle(a \sigmap + \adagger \sigmam)\right),
\end{equation}
where we define an effective interaction strength
$    \angle= \sqrt{\gamma\dt} $
where $\sqrt{\gamma/\Delta t}$ is the coarse-grained coupling strength between the qubit and cavity~\cite{Fischer2018May, Gross2018Feb, Ciccarello2017Dec}.
Since the qubit is always in the ground state before interaction, this excitation exchange results in photons being transferred out of the cavity, corresponding to an effective loss rate $\gamma = \angle ^2/\dt $. The parameter $\gamma$ would be the loss rate for the open systems model in the continuous limit, and it corresponds to a measurement rate in our protocol, akin to a decay into the measurement apparatus instead of an uncontrolled environment.
Each interaction via the unitary~\eqref{eq:unitary} can also be regarded as a partial SWAP for small $\angle$, since the operation corresponds to an iSWAP for $\angle=\swapangle=\pi/2$.


During the interaction-measurement sequence, the cavity state will evolve stochastically as a quantum trajectory conditioned on each choice of the measurement basis and qubit measurement result. Measuring in the $x$ or $y$ basis, or alternating between these, gives rise to diffusive trajectories of the state corresponding to homodyne or heterodyne detection, respectively~\cite{Gambetta2005Sep, Daryanoosh2022Aug}, while measuring in the $z$ basis gives rise to quantum jump dynamics corresponding to photodetection.
We utilize the model to show that the \emph{measurement record}, by choice of qubit observable, provides the same statistics as quadrature measurements of the cavity.

%

 \section{Qubitdyne: digital homo- and heterodyne detection}
An ideal homodyne detector measures the \emph{generalized quadrature}
\begin{equation}\label{eq:generalized_quadratures}
\begin{split}
    x_\phase &= \frac{1}{\sqrt{2}}(\adagger\expup{\imi\phase} + a \expup{-\imi \phase}), \\
\end{split}
\end{equation}
which reduces to the ordinary $x$- and $p$-quadratures for $\phase=0$ and $\phase=\pi/2$, respectively. 
Just as the measured quadrature is chosen by the phase angle of a local oscillator in a regular homodyne measurement, our measurement selects a quadrature by choosing a measurement axis in the $xy$ plane for the qubit. In practice, such a measurement can be performed as a $\pi /2$ rotation along the desired axis, followed by a measurement in the computational basis.
The corresponding expectation value is a measure of the cavity field quadrature given by Eq.~\eqref{eq:generalized_quadratures}, since there is a direct relation between qubit and cavity expectations (see Supplemental Material~\cite{supplemental})
\begin{equation}\label{eq:operator_correspondence}
    \braket{\sigmam} = - \imi \sqrt{\gamma\dt} \braket{a}.
\end{equation}

Expectation values are the average over an ensemble of quantum trajectories. To determine the full probability distribution of measurement outcomes, we consider individual trajectories.
If $X_n$ is the random variable corresponding to qubit measurement outcome $\pm 1$ at step $n$, the measurement value of one trajectory with $\nmeas$ qubit measurements is given by the random variable
\begin{equation}\label{eq:meas_J}
    J_{\rm hom} = \sum_n^{\nmeas}  f(t_n) X_n,
\end{equation}
%
where the digital measurement result is weighted by the function
\begin{equation}\label{eq:exp_weights}
    f(t_n)=\sqrt{\gamma\dt/2}\exp(-\gamma t_n /2),
\end{equation}
at time step $t_n=n\dt$. The exponential form can be understood from a comparison to ordinary homodyne measurement. For a field to be measured with high efficiency, it needs to be mode-matched~\cite{Ou1995Oct}. Decay of the cavity field through repeated interactions corresponds to the field leaking into a fictitious waveguide (chain of qubits) with  rate $\gamma$, where Eq.~\eqref{eq:exp_weights} corresponds to the probability amplitude used for temporal mode matching. 
Below, we simulate $\ntrajs$ realizations of the stochastic process~\eqref{eq:meas_J} to obtain statistics.

\begin{figure}[h!]
    \centering
    \includegraphics[width=\columnwidth]{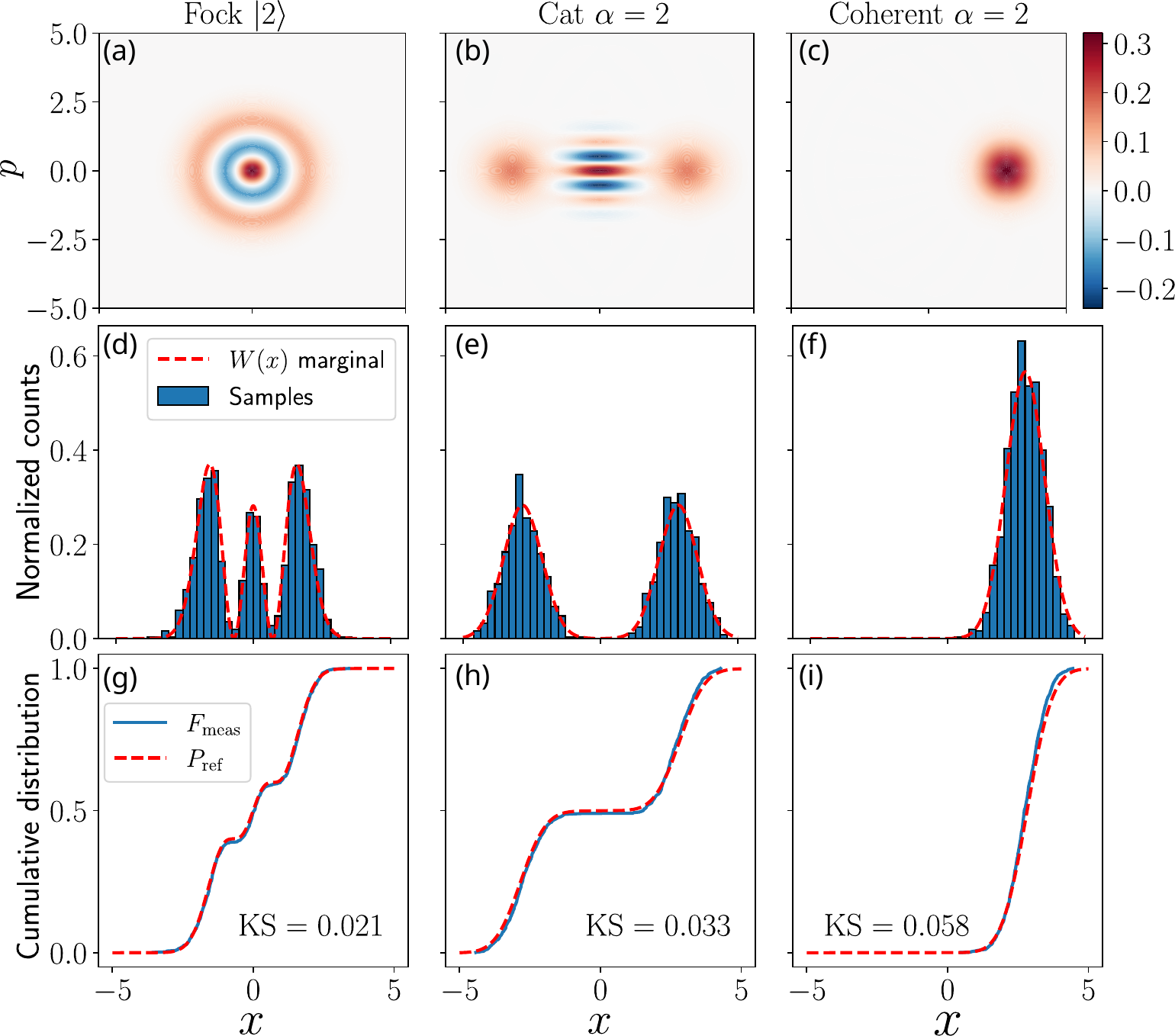}
    \caption{Simulated qubitdyne measurements corresponding to a homodyne record. Each column corresponds to a unique state.
    Top row: Ideal Wigner functions.
    Middle row: Histograms from 1000 measurement rounds, compared to the ideal Wigner marginals (dashed lines).
    Bottom row: Simulated cumulative distribution functions $F_\text{meas}$ (solid lines) and ideal distributions $P_\text{ref}$ (dashed lines). The distance between the two distributions is quantified by the Kolmogorov-Smirnov (KS) statistic. Simulations used interaction strength $\angle = 0.1\swapangle$ and $\nmeas=200$ qubit measurements. }
    \label{fig:homodyne_wigner_hist}
\end{figure}

The probability distribution for homodyne detection is the marginal distribution of the Wigner function along the measured quadrature~\cite{Leonhardt1997-kr}, and we show that the values $J_{\rm hom}$ are sampled from this distribution. As an example, we show simulated measurement statistics for three different states, Fock $\ket 2$, cat $(\ket{\alpha} + \ket{-\alpha})/\sqrt{2}$ and coherent $\ket \alpha$ states with $\alpha=2$, whose Wigner functions are displayed in Figs.~\figpanel{fig:homodyne_wigner_hist}{a-c}.
Normalized histograms with values calculated as Eq.~\eqref{eq:meas_J} from $\ntrajs=1000$ simulated measurement rounds with $\nmeas=200$ qubit measurements each are shown in Figs.~\figpanel{fig:homodyne_wigner_hist}{d-f}.
As a quantitative measure of how well our digitized homodyne measurement corresponds to an ideal measurement, we use the Kolmogorov-Smirnov (KS) statistic~\cite{conover_1998, Massey1951Mar}. 
The KS statistic measures the largest vertical distance between the empirical and reference cumulative distribution functions $F_\text{meas}(x)$ and $P_\text{ref}(x)$:
 \begin{equation}
     \mathrm{KS} = \max _x |F_\text{meas}(x) - P_\text{ref}(x)|.
 \end{equation}
The empirically sampled distributions and the ideal distributions calculated from the Wigner marginals, along with the KS statistics, are displayed in Figs.~\figpanel{fig:homodyne_wigner_hist}{g-h}. We can also use the fidelity of a state reconstruction as a proxy for the statistical accuracy of the data. Using the reconstruction method from Ref.~\cite{Strandberg2022Oct}, a fidelity of 0.99 is obtained for all three states with simulated measurements of 10 quadrature angles.

\begin{figure}[h!]
    \centering
    \includegraphics[width=\columnwidth]{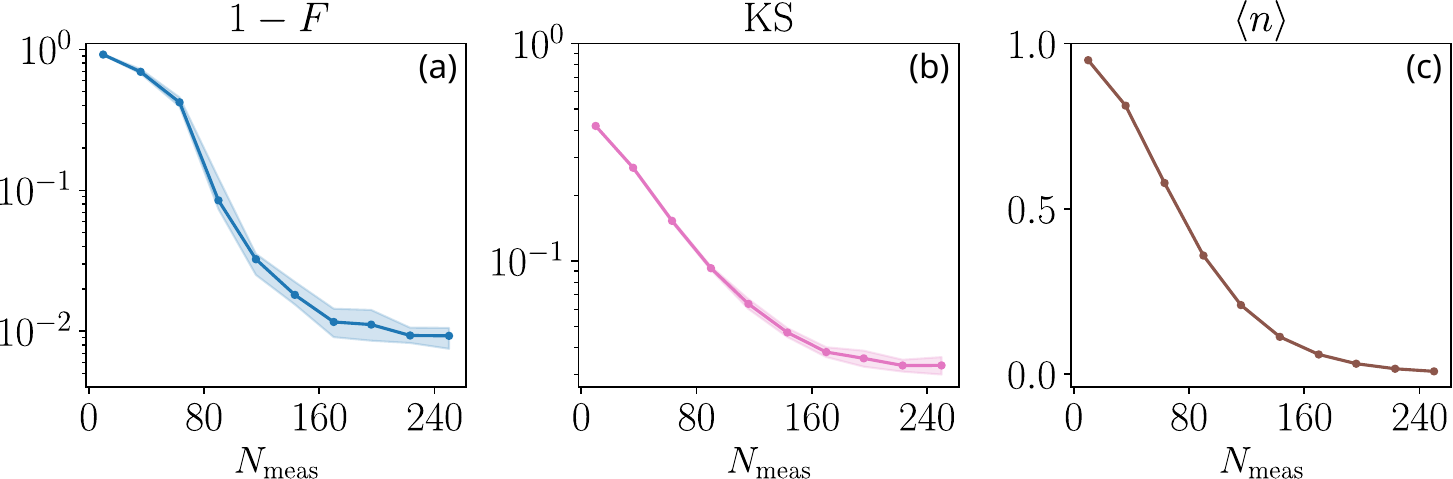}
    \caption{Measures of statistical accuracy and the final cavity population as a function of the number of qubit measurements $\nmeas$, for a cat state of amplitude $\alpha=2$. (a) Infidelity between the ideal state and the state reconstruction from sampled data. The shaded region indicates the standard deviation of 10 different tomography rounds. (b) Kolmogorov-Smirnov statistic of the sampled data. (c) Cavity population at the end of a measurement round.
    }
    \label{fig:cat_loop_nmeas}
\end{figure}

There are two key criteria that need to be met to produce accurate measurement statistics: (i) the qubit excitation probability must be small, and (ii) the cavity must be almost empty by the final measurement.
The probability to excite the qubit during an interaction is
\begin{equation}
    p_e = \gamma \dt \braket{\adagger a} = \angle^2 \braket{\adagger a},
\end{equation}
depending not only on the interaction strength but also on the average cavity photon number $\braket{\adagger a}$. For any given cavity state, the effective interaction strength must be chosen such that the condition
\begin{equation}\label{eq:conition:pe}
    p_e \ll 1,
\end{equation}
is fulfilled.
%
Additionally, the state must be sufficiently extracted from the cavity to obtain complete information. This requirement means that for a given interaction strength, a particular minimum number of qubit measurements $\nmeas$ is needed, such that $\nmeas \angle^2 \gg 1$.
Using the same cat state as before, again with interaction $\angle = 0.1\swapangle$ and $\ntrajs=1000$, Fig.~\ref{fig:cat_loop_nmeas} shows the infidelity, KS statistic, and final cavity population for different values of $\nmeas$. It can be seen that accurate statistics are only obtained when most of the state has been extracted at the end of a measurement round. In this example, a fidelity of 0.99 is first obtained when the cavity field has reached around \SI{94}{\%} vacuum. 



\begin{figure}[h!]
    \centering
    \includegraphics[width=\columnwidth]{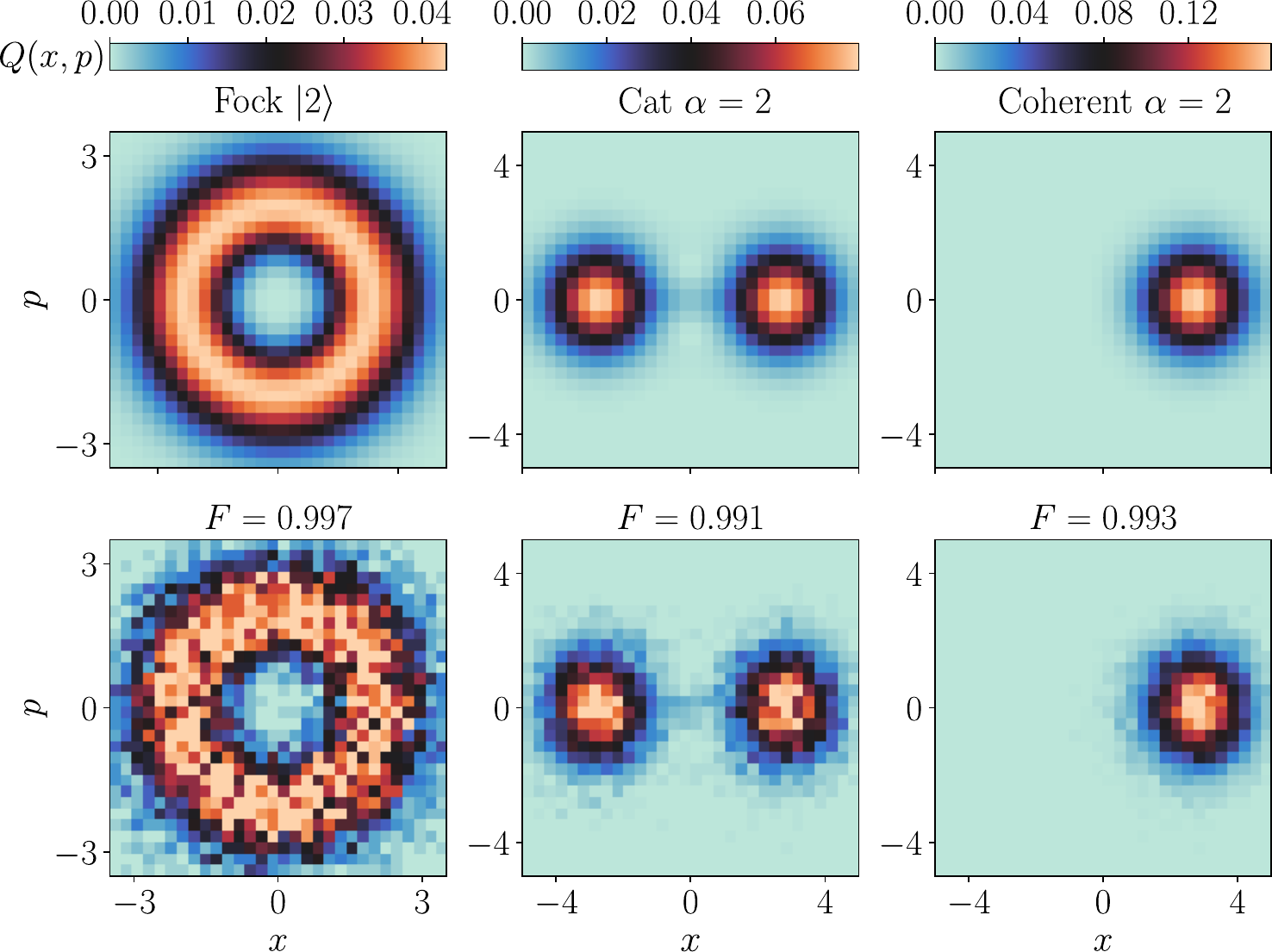}
    \caption{Qubitdyne measurements corresponding to heterodyne detection. Top row: Ideal discretized $Q$-functions for three different states. Bottom: Heterodyne histograms obtained from alternating $\sigma_x$ and $\sigma_y$ measurements, using an interaction strength $\angle=0.1\swapangle$ and $\nmeas=300$ qubit measurements each round. Tomographic fidelity $F> 0.99$ for all three states.
    }
    \label{fig:heterodyne_hists}
\end{figure}

Next, we present heterodyne detection, which measures two orthogonal quadratures simultaneously at the cost of additional measurement noise originating from Heisenberg's uncertainty principle.
Heterodyne measurement statistics is obtained by interleaving $\sigma_y$ and $\sigma_x$ measurements. Referring back to Fig.~\figpanel{fig:system}{a}, this scheme is equivalent to dividing the qubits into two separate streams, alternating $\sigma_y$-measurements on one stream and $\sigma_x$-measurements on the other.
The value from one measurement round with a total of $\nmeas$ qubit measurements is therefore given by
\begin{equation}
    J_{\rm het} = \sum_{n=1}^{\nmeas-1} 2\left[f(t_n)Y_n + \imi f(t_{n+1})X_{n+1} \right],
\end{equation}
with the weight function~\eqref{eq:exp_weights}, $X_n$ being the outcome of $\sigma_x$-measurements, and $Y_n$ of $\sigma_y$-measurements.
Two-dimensional histograms of $\ntrajs=10\, 000$ measurement rounds with $\nmeas=300$ qubit measurements are shown in the bottom row of Fig.~\ref{fig:heterodyne_hists} for the state $\ket 2$, cat $(\ket{\alpha} + \ket{-\alpha})/\sqrt{2}$ and coherent state $\ket \alpha$. These histograms can be compared to the top row of Fig.~\ref{fig:heterodyne_hists} which shows the state $Q$-functions, corresponding to ideal heterodyne measurements. The fidelity of state reconstructions with the sampled data reaches at least 0.99.


%
%



\section{Effect of finite cavity lifetime}

In the optimal scenario without dissipation, measurements approach ideal statistics in the continuous measurement limit, which is attained by reducing $\angle$ and increasing $\nmeas$. However, in a realistic setting, the cavity dissipation rate $\kappa$ sets a limit on how many measurements can be made before the cavity state has leaked into the environment. Hence, there will be an optimal interaction strength, depending on the cavity decay rate and initial state.

We now show that the qubitdyne scheme is expected to work with high fidelity for multiphoton states with an average of up to six photons in a cavity with a ratio 1:500 between the duration $T$ of one measurement step and the cavity lifetime $T_1$. This ratio is accessible, for instance, in superconducting circuits, assuming a lifetime $T_1 = 1/\kappa = \SI{500}{\micro\second}$ for a 3D microwave cavity~\cite{Kudra2022Jul} and a total measurement time $T=\SI{1}{\micro\second}$, the latter encompassing the duration of qubit-cavity interaction \cite{Kudra2022Dec}, qubit measurement~\cite{walter2017rapid}, and qubit reset~\cite{magnard2018fast}.



%

\begin{figure}[h!]
    \centering
    \includegraphics[width=\columnwidth]{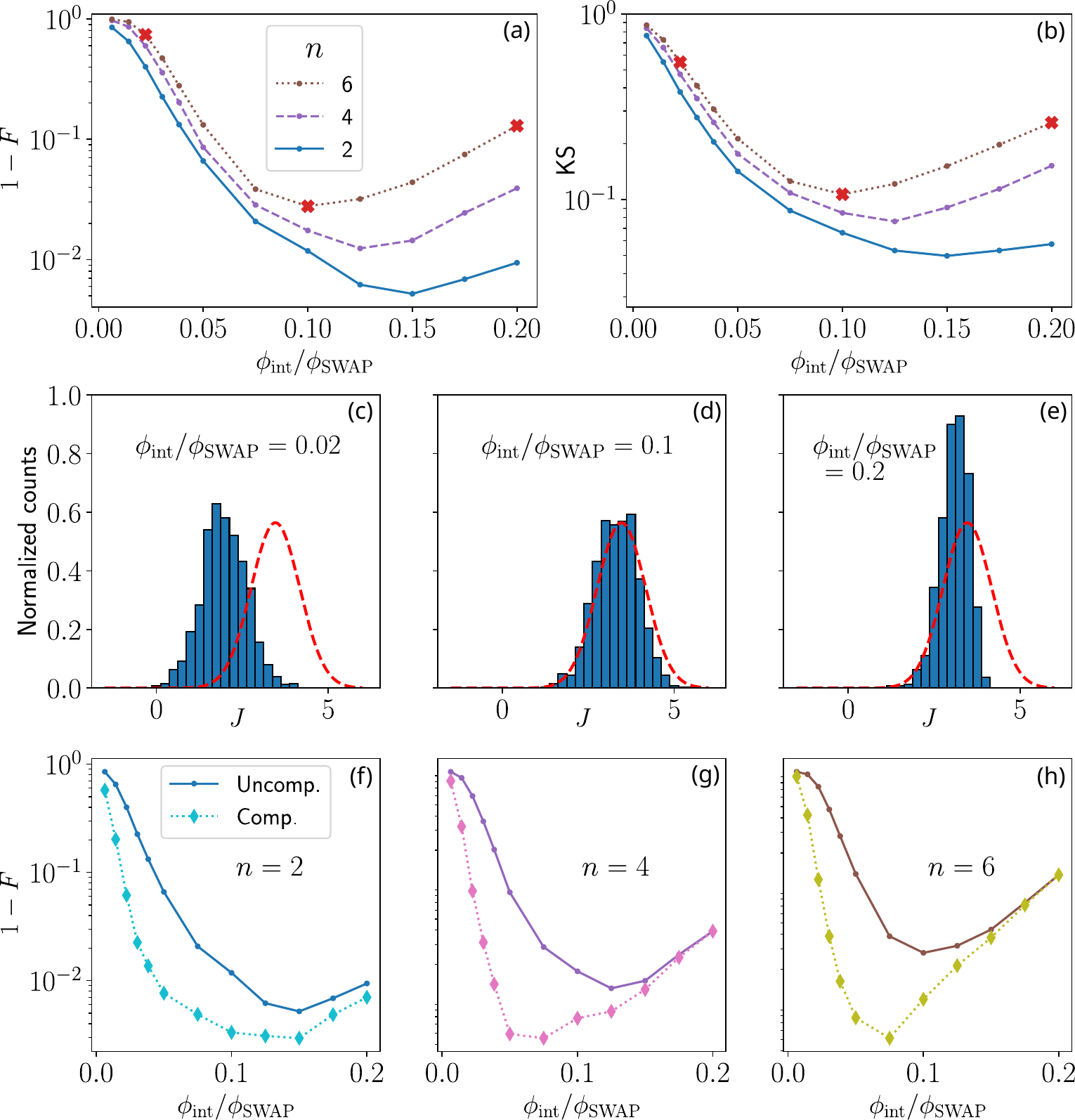}
    \caption{Homodyne measurement statistics with external cavity dissipation for coherent states with average photon numbers $2,4$ and $6$. (a) Tomographic infidelity as a function of interaction strength $\angle$. (b) KS statistic as a function of $\angle$. The  markers on the $n=6$ line indicate interaction strengths for which histograms are visualized in Figs. (c-e) Simulated homodyne histograms and corresponding ideal distributions (dashed lines). (c) This interaction strength is too weak and the histogram is shifted towards vacuum. (d) The histogram matches the expected distribution, this is the optimal interaction strength. The corresponding efficiency at this point is $\eta=0.925$. (e) The interaction is too strong, leading to a distorted histogram. (f-h) Infidelities from Fig. (a) (solid lines) and noise-compensated infidelities (dashed lines).
    }
    \label{fig:loop_theta}
\end{figure}

The interplay between cavity decay and measurement strength is visualized in Fig.~\ref{fig:loop_theta}. Figures~\figpanel{fig:loop_theta}{a-b} show the infidelity and KS statistic for coherent states with different photon numbers as a function of interaction strength. For each $\angle$, the number of measurements was set such that the cavity was \SI{95}{\%} vacuum at the end of each measurement round.
Histograms of the simulated measurement statistics at three different interaction strengths for $n=6$ photons can be seen in Figs.~\figpanel{fig:loop_theta}{c-d}, illustrating three different regimes. First, in panel (c), the interaction is very weak and the field is mostly decaying into the environment, leading to the measured distribution being mixed with vacuum. In panel (d), an optimal interaction strength is reached. Finally, for a stronger interaction in panel (e), the distribution is distorted because condition~\eqref{eq:conition:pe} is violated.
%
%
%
%
This condition is necessary for the protocol to be valid, but the effect of a finite cavity lifetime as shown in Fig.~\figpanel{fig:loop_theta}{c} simply corresponds to a nonideal collection efficiency.
With a measurement rate $\gamma$ and radiative decay rate $\kappa$, the photon collection efficiency is
\begin{equation}
      \eta = \frac{\gamma}{\gamma + \frac{T}{\dt}\kappa}.
\end{equation}
The ratio $T/\dt$ appears since intrinsic loss with rate $\kappa$ occurs throughout the entire protocol, but the measurement rate  $\gamma$ is only activated during the shorter interaction times $\dt$.
Generally, a reduced efficiency $\eta < 1$ degrades the measured distributions. However, for the purpose of tomographic measurements, it can be compensated for~\cite{D'Ariano1998Apr} to obtain a reliable state reconstruction. The effect of this compensation is shown in Figs.~\figpanel{fig:loop_theta}{f-h}, where infidelity is reduced in the regime of small $\angle$.

The maximal collection efficiency for a coherent state with average photon number $n=6$ is $\eta=0.925$. Assuming an efficiency $\eta_q=0.98$ from qubit readout error, the total detection efficiency is $\eta_{\rm det}=\eta_q \eta = 0.91$. This is more than twice the detection efficiency when releasing a multiphoton state of similar size~\cite{Pfaff2017Sep}.
%

\section{Enhanced readout speed}
As shown in Fig.~\ref{fig:cat_loop_nmeas}, most photons must be removed from the cavity via the qubit to obtain correct information about the state. The constant coupling $\angle$ gives rise to an exponential cavity decay, but the number of measurements needed to sufficiently empty the cavity could still be large, particularly for states with high photon numbers. As seen above, this slow execution of the protocol leads to reduced measurement fidelity in the presence of loss. A way to alleviate this problem is to reduce the number of needed measurement by successively increasing the qubit-cavity coupling rate, which allows the cavity to be emptied faster while still keeping the qubit excitation probability low. To obtain accurate measurement statistics for a time-dependent coupling $\angle(t)$, we find that the the appropriate weight function $f(t)$ has the shape corresponding to the temporal envelope of a single-photon wavepacket emitted via such modulation~\cite{Gough2015Dec, Li2022Sep, Gheri1998Jun}. The expression can be obtained by solving the quantum Langevin equation along with using the input-output relation, and the resulting expression is given by
%
\begin{equation}
    f(t) = \frac{\angle(t)}{\sqrt{2}}\exp\left(-\frac{1}{2\dt}\int_0^t\angle(t)^2 \dif t'\right).
\end{equation}
As we show in the Supplementary Materials~\cite{supplemental}, the use of a time-dependent coupling strength can reduce the number of measurements by a factor 2, while preserving the measurement fidelity.
%

\section{Discussion}

The measurement scheme presented in this letter opens the door for quantum information processing protocols that require homo- or heterodyne detection of confined cavity modes.
The scheme only requires a beamsplitter or SWAP interaction between a two-level system and the bosonic mode of interest, and the ability to repeatedly measure the qubit. This simplicity makes it applicable on a large variety of platforms.

Under realistic conditions, we expect the qubitdyne protocol to perform better than the release of a microwave mode into a transmission line, followed by detection with a nearly-quantum limited parametric amplifier. This because the latter measurement is limited by the finite efficiency of the amplification chain used to measure the field quadratures~\cite{Pfaff2017Sep}, while highly accurate qubit readout is possible even without a quantum-limited amplifier~\cite{Chen2023Mar}.

As an alternative version of qubitdyne, one may consider to implement a phase estimation protocol~\cite{nielsen_quantum_2010} of the displacement operator, which amounts to a modular quadrature coordinate measurement~\cite{Tehral2016Jan} (see Supplemental Material). However, we expect the phase estimation approach to be more sensitive to loss channels than regular qubitdyne because the amount of energy present in the cavity increases with each phase-estimation round, while the cavity is emptied in the presented qubitdyne protocol.

Among possible applications of qubitdyne, we envisage efficient boson sampling verification~\cite{Chabaud2021Nov} and quantum state certification~\cite{Chabaud2021Jun}.
The code used to perform numerical simulations in this work is available at~\cite{code}.

\section*{Acknowledgements}

SG is thankful to Giulia Ferrini, Alessandro Ferraro, and Ulysse Chabaud for useful discussions which initiated this work. IS gives thanks to Maryam Khanahmadi for important insights regarding the time-dependent coupling.
Simulations were done using QuTiP~\cite{Johansson2013Apr}.
This work is supported by the Knut and Alice Wallenberg foundation via the Wallenberg Centre for Quantum Technology (WACQT). IS acknowledges financial support from the European Union via \mbox{Grant No.\ 101057977 SPECTRUM}. BR ackowledges funding from the Canada First Research Excellence Fund, the Natural Sciences and Engineering Research Council of Canada (NSERC) as well as the  Fonds de Recherche du Québec, Nature et Technologie (FRQNT).

\bibliography{references}%

\newpage

\onecolumngrid

\include{supplemental}
\end{document}

%% file: supplemental.tex
\title{Supplemental Material for\\``Digital homodyne and heterodyne detection for stationary bosonic modes''}

\author{Ingrid Strandberg}%
\affiliation{Department of Microtechnology and Nanoscience MC2, Chalmers University of Technology, SE-412 96
G\"oteborg, Sweden}

\author{Axel Eriksson}
\affiliation{Department of Microtechnology and Nanoscience MC2, Chalmers University of Technology, SE-412 96
G\"oteborg, Sweden}

\author{Baptiste Royer}
\affiliation{Département de Physique and Institut Quantique, Université de Sherbrooke, Sherbrooke J1K 2R1 QC, Canada}

\author{Mikael Kervinen}
\affiliation{Department of Microtechnology and Nanoscience MC2, Chalmers University of Technology, SE-412 96
G\"oteborg, Sweden}
\affiliation{VTT Technical Research Centre of Finland Ltd. Tietotie 3, Espoo 02150, Finland}

\author{Simone Gasparinetti}
\affiliation{Department of Microtechnology and Nanoscience MC2, Chalmers University of Technology, SE-412 96
G\"oteborg, Sweden}

\maketitle

\renewcommand{\theequation}{S\arabic{equation}}
\renewcommand{\thefigure}{S\arabic{figure}}
\renewcommand{\thetable}{S\arabic{table}}

\tableofcontents

\section{State evolution}

The system evolution can be derived within the formalism of quantum filtering ~\cite{Bouten2009May, Gough2004Sep} and quantum stochastic calculus~\cite{Barchielli1996Feb, Pellegrini2008Nov}, leading directly to a stochastic differential equation describing the system dynamics. Here we use the repeated interactions formalism to show that repeated indirect measurements lead to the same results.

With the cavity initially in state $\psin{0}$ and the qubit always in the ground state $\ket g$ before each interaction, the initial system state is $\ket{\phi_0}=\psin{0}\otimes\ket g$.
After the $n$th interaction, the state is
\begin{equation}\label{eq:evolved_state_U}
    \ket{\phi_{n+1}}=U(\psin{n}\otimes\ket g)
\end{equation}
and after the projective qubit measurement it is again a product state $\psin{n+1}\otimes\ket i$
with the cavity in the state
\begin{equation}
    \psin{n+1} = \frac{K_{i_{n}} \ket{\psi_{n}}}{\braket{\psi_{n}|K_{i_{n}}^\dagger K_{i_{n}} |\psi_{n}}}
\end{equation}
where $i_n$ is the measurement result at the $n$th step. For $\ket i=\{\ket g,\ket e\}$ or any other orthonormal basis, the \emph{Kraus operators} are defined by the partial inner product $K_i = \braket{i|U|g}$ and act as~\cite{Bauer2011Oct}
\begin{equation}
    U\ket{\phi_n}=\sum_i \left( K_i \psin{n} \right) \otimes \ket i.
\end{equation}
The probability of each measurement result $i$ at step $n$ is
\begin{equation}\label{eq:kraus:_prob}
    p_i = \braket{\psi_n|K_i^\dagger K_i |\psi_n}.
\end{equation}
As such, it is clear that the probability to get the qubit measurement result $i$ is conditioned on the cavity state $\ket{\psi_n}$. In turn, the sequence of measurements induces changes in the system due to quantum back-action.




The state evolution can be calculated explicitly.
In the weak coupling regime and for short interaction times such that
\begin{equation}\label{eq:criteria2}
    \sqrt{\gamma\dt} \ll 1,
\end{equation}
we Taylor expand the unitary Eq.~(1) to order $\gamma\dt$:
\begin{equation}
    U = \mathds{1}  -\imi\sqrt{\gamma\dt}(a\sigmap + \adagger\sigmam) - \frac{\gamma\dt}{2}(a\sigmap + \adagger\sigmam)^2.
\end{equation}
Because of the identity $\sigmap^2=\sigmam^2=0$ and the fact that the qubit is initialized to its ground state before the interaction, this can be simplified to
\begin{equation}
    U = \mathds{1} - \imi\sqrt{\gamma\dt}a\sigmap - \frac{\gamma\dt}{2}\adagger a\sigmam\sigmap.
\end{equation}
Explicitly calculating the evolved state~\eqref{eq:evolved_state_U}, we get
\begin{equation}\label{eq:phi1}
    \ket{\phi_{n+1}} = [ 1 - \frac{\gamma\dt}{2}\adagger a] \psin{n}\otimes\ket g - \imi \sqrt{\gamma\dt}a\psin{n}\otimes\ket e.
\end{equation}
From here on we suppress the subscript $n$ for notational simplicity.
The expectation value $\braket{\sigma_x}$ for the evolved state~\eqref{eq:phi1} can be calculated using $\sigma_x = \ketbra{e}{g} + \ketbra{g}{e}$ to be
\begin{equation}\label{eq:exp_sx2}
    \braket{\phi|\sigma_x|\phi} = \imi\sqrt{\gamma\dt}\braket{\psi|\adagger -a |\psi} =  \sqrt{2\gamma\dt}\braket{p}
\end{equation}
to order $\gamma\dt$, with the cavity $p$-quadrature defined as
\begin{equation}\label{eq:p_quad}
    p = \imi\frac{\adagger-a}{\sqrt{2}}.
\end{equation}
We can see that the qubit expectation value $\braket{\sigma_x}$ is proportional to the expectation of the cavity $p$-quadrature.
A corresponding calculation gives the relation \mbox{$\braket{\sigma_y}=-\sqrt{2\gamma\dt}\braket{x}$}.

These relations are decided by the sign-choice in the interaction Hamiltonian. Here we considered the unitary generated by $\hint = \sqrt{\gamma/\Delta t}(a\sigmap + \adagger\sigmam)$, but the opposite, i.e., $\braket{\sigma_x}$ corresponding to the $x$-quadrature, would occur for $\hint = \imi\sqrt{\gamma/\Delta t}(a\sigmap - \adagger\sigmam)$. In practice, any quadrature can be measured by rotating the qubit before measurement in the computational basis.

\section{Correspondence to stochastic homodyne signal}

Previous work~\cite{Gross2018Feb} show that the equation of motion due to repeated interactions corresponds to a stochastic master equation in the continuous limit. Here we extend this result to show that the indirect measurement signal also corresponds to the correct stochastic equation.
To show the correspondence of the qubit measurement outcome to the homodyne signal for the usual stochastic Schrödinger or master equation, we utilize the \emph{Doob decomposition}~\cite{klebaner2005introduction,Doob1990-tf} of the counting process
\begin{equation}\label{eq:counting_process}
    N_n = \sum_{k=1}^n X_k.
\end{equation}
which is the sum of random variables corresponding to qubit measurement outcomes $X_k=\pm 1$. Similarly as in Refs.~\cite{Bauer2013May, Bauer2014Sep}, we decompose Eq.~\eqref{eq:counting_process} into two unique terms:
\begin{equation}
N_n = A_n + M_n.
\end{equation}
into a predictable process $A_n$:
\begin{equation}\label{eq:predictable}
    A_n = \sum_{k=1}^n (E[N_k|\filter{k-1}] - N_{k-1}),
\end{equation}
And a martingale
\begin{equation}\label{eq:martingale}
    M_n = \sum_{k=1}^n (N_k - E[N_k|\filter{k-1}])
\end{equation}
for $N_0$ = 0. The operation $E[\,\cdot\,|\,\cdot\,]$ denotes the conditional expectation value, $\filter{k}$ is called a \emph{filtration} in classical stochastic calculus, and it encapsulates information of outcomes $\{N_i\}_{i<k}$.

In Eq.~\eqref{eq:predictable} the expectation values of $N_k$ are conditioned on past measurement results, and the fact that the process is predictable means that the value at step $n$ is determined by the value at step $n-1$. More technically, it means $A_n$ is \mbox{$\filter{n-1}$-measurable} for all $n$. When we go to the continuous-time limit it will converge to the drift term in the stochastic differential equation. A martingale process has the property that at any time, the conditional expectation of the next value is equal to the present value~\cite{klebaner2005introduction,Calin2015-ea}. As such, we can see from definition~\eqref{eq:martingale} that it has zero mean. It will converge to the Brownian noise term in the stochastic equation, as indeed, standard Brownian motion is a zero-mean martingale with variance 1.


We now calculate the increment
\begin{equation}\label{eq:DeltaA}
   \Delta A = A_{n+1} - A_n = E[N_{n+1} |\filter{n}] - N_n.
\end{equation}
The variable $N_n$ is $\filter{n}$-measurable, meaning $N_n = E[N_n|\filter{n}]$, and due to linearity of the expectation we can rewrite~\eqref{eq:DeltaA} as
\begin{equation}
    E[N_{n+1} |\filter{n}] - N_n = E[N_{n+1} - N_n|\filter{n}]  = E[\Delta N|\filter{n}].
\end{equation}
From Eq.~\eqref{eq:counting_process} we can see that
\begin{equation}\label{eq:NtoX}
    N_{n+1} = N_n + X_{n+1} \implies \Delta N = X_{n+1}.
\end{equation}
Using this, we end up with
\begin{equation}
    \Delta A = E[X_{n+1}|\filter{n}].
\end{equation}
This is the expectation value of the measurement outcome at step $n+1$ conditioned on past measurement results up to and including time step $n$. Since the stochastic process is Markovian, the direct dependence is only on the previous step $n$ and we can conclude that
\begin{equation}
     E[X_{n+1}|\filter{n}] = \braket{\psi_{n}|\sigma_x|\psi_{n}}.
\end{equation}
where from now on we only write $\braket{\sigma_x}$ for simplicity. Using $\Delta N = \Delta A + \Delta M$ and Eq.~\eqref{eq:NtoX}, we have for each measurement the random variable
\begin{equation}\label{eq:meas_random}
    X_{n+1} = \braket{\sigma_x} + \Delta M .
\end{equation}
%

The repeated measurements gives rise to open system dynamics with decay rate $\gamma$. Analogously to temporal mode matching in ordinary homodyne detection, we scale Eq.~\eqref{eq:meas_random} with the exponential filter function
\begin{equation}\label{eq:weight_function}
    f(t_n) = \frac{1}{\sqrt{2}}\sqrt{\gamma\dt}\expup{-\gamma t_n /2}.
\end{equation}
It can be noted that in continuous measurements, a mode function is normalized to $\int |f(t)|^2 \dif t=1$ to construct the operator $a_f = \int a(t)f(t)\dif t$ and its conjugate that fulfill the bosonic commutation relation. Here we instead combine the mode function with a quantity corresponding to a generalized quadrature, and have normalization $\sum_n |f(t_n)|^2=1/2$ such that the minimum quadrature variance is 1/2.

Defining the stochastic increment $\Delta W = \sqrt{\dt}\Delta M$ and also inserting the expectation~\eqref{eq:exp_sx2}, the measurement value is
\begin{equation}\label{eq:meas_X}
    j_n = f(t_n) X_n = (\sqrt{\gamma}\braket{p}\dt + \frac{1}{\sqrt{2}} \Delta W ) \sqrt{\gamma}\expup{-\gamma t_n/2},
\end{equation}
The final result~\eqref{eq:meas_X} has the expected form: a deterministic term proportional to the quantum conditioned average of the quadrature, plus the stochastic Wiener increment with zero mean and variance $\dt$~\cite{Wiseman1993Jan} (since $\Delta M$ has variance 1). Adding the values $j_n$ for all $n$ corresponds to the integrated measurement signal of a homodyne measurement. Eq.~\eqref{eq:meas_X} can be related to the well-known stochastic equation for homodyne detection with measurement signal~\cite{Gardiner, Wiseman2009-ti}
\begin{equation}\label{eq:dj}
   \dif j=\sqrt{\gamma}\braket{p}\dif t + \frac{1}{\sqrt{2}}\dif W,
\end{equation}
where the factor $1/\sqrt{2}$ for the stochastic increment ensures a normalization where a measurement of vacuum gives a Gaussian distribution with variance 1/2.
Mode-matching to the standard exponential decay of a single excitation is obtained by multiplying the signal Eq.~\eqref{eq:dj} with the temporal mode function~\cite{Milburn2015Aug,Pozza2015Jan} \mbox{$f(t)=\sqrt{\gamma}\exp(-\gamma t /2)$}, giving an exact correspondence to Eq.~\eqref{eq:meas_X}.

\section{Filter function relation to quadrature definition}

In quantum optics there are different conventions for the quadrature scaling when defined in terms of creation and annihilation operators. Which scaling is obtained by qubitdyne measurements can be chosen by setting a constant in the filter function, as listed in Table~\ref{tab:quadrature_defs} for homodyne detection. For instance, the factor of $1/\sqrt{2}$ in Eq.~\eqref{eq:weight_function} corresponds to the quadrature definition
\begin{equation}
    x = \frac{a + \adagger}{\sqrt{2}},
\end{equation}
which has a vacuum fluctuation variance $(\Delta x)^2_\text{vac}=1/2$.

The chosen quadrature definition is obtained from heterodyne measurements by setting its filter function scaling to twice the homodyne value in Table~\ref{tab:quadrature_defs}, i.e. \mbox{$f(t_n) = 2c \times \sqrt{\gamma\dt}\expup{-\gamma t_n /2}$}. This scaling gives a heterodyne vacuum variance that is twice that of homodyne measurement, which is the expected result due to added noise arising as a consequence of joint detection of non-commuting quadratures~\cite{Leonhardt1997-kr}.

\begin{table}[h!]
    \centering
    \begin{tabular}{lccc}
        &$f(t_n) = c \times \sqrt{\gamma\dt}\expup{-\gamma t_n /2} $  & $x$ & $(\Delta x)^2_\text{vac}$  \\
         \hline
         &$c=1/2$ & $\dfrac{a + \adagger}{2}$ & 1/4 \\
        &$c=1/\sqrt{2}$ & $\dfrac{a + \adagger}{\sqrt{2}}$  & 1/2 \\
         &$c=1$ & $a + \adagger$  &  1\\
    \end{tabular}
    \caption{Different quadrature definitions can be chosen by scaling the homodyne filter function $f(t_n)$. The conventions lead to different values of minimum vacuum fluctuations $(\Delta x)^2_\text{vac}$. }
    \label{tab:quadrature_defs}
\end{table}

\section{Time-dependent coupling}

To obtain the correct filter function to weigh the qubit measurement results with, we solve the quantum Langevin equation for the time-dependent cavity field operator $a$, where the time argument is omitted for brevity. The equation is~\cite{Gardiner}
\begin{equation}
    \dod{a}{t} = - \frac{\gamma}{2}a - \sqrt{\gamma}a_\text{in},
\end{equation}
for a cavity with decay rate $\gamma$ and an input field $a_\text{in}$. The input-output relation is
\begin{equation}
    a_\text{out} = a_\text{in} + \sqrt{\gamma}a.
\end{equation}
With vacuum input, the equations for the corresponding field amplitudes become
\begin{equation}
    \dod{\alpha}{t} = - \frac{\gamma}{2}\alpha, \quad  \alpha_\text{out} =\sqrt{\gamma}\alpha.
\end{equation}
If the decay rate is time dependent, $\gamma= \gamma(t)$, the solution of the output amplitude is~\cite{Gheri1998Jun}
\begin{equation}
    \alpha_\text{out}(t) = \sqrt{\gamma(t)} \expup{-\frac{1}{2}\int_0^t \gamma(t')\dif t'}.
\end{equation}
The cavity photon number corresponds to the intensity of the field, which gives the solution
\begin{equation}
    n(t) = |\alpha(t)|^2 = n_0 \expup{-\int_0^t \gamma(t')\dif t'},
\end{equation}
fulfilling the boundary condition $n(0)=n_0$ being the initial photon number.
The qubit excitation probability is
\begin{equation}
    p_e(t) = \gamma(t) \dt n(t) = \gamma(t) \dt n_0 \expup{-\int_0^t \gamma(t')\dif t'}.
\end{equation}
The properly "mode-matched" filter function is related to this probability as
\begin{equation}
     f(t) = \sqrt{p_e(t)/2n_0},
\end{equation}
giving the final result
\begin{equation}
    f(t) = \sqrt{\gamma(t)\dt/2}\expup{-\frac{1}{2}\int_0^t \gamma(t')\dif t'}.
\end{equation}

\section{Alternative derivation of time-dependent coupling with loss}

To obtain the correct filter function to weigh the qubit measurement results with, we solve the quantum Langevin equation for the time-dependent cavity field operator $a$, where the time argument is omitted for brevity. It is implicitly assumed that the duration of one measurement only consists of the interaction time $\dt$, but this can simply be generalized by adding time-scaling factors as in the main text. The equation to solve is~\cite{Gardiner}
\begin{equation}
    \dod{a}{t} = - \frac{(\gamma + \kappa)}{2}a - \sqrt{\gamma}a_\text{in}  - \sqrt{\kappa}b_\text{in},
\end{equation}
for a cavity with an engineered decay rate $\gamma$ which can vary in time, an external decay rate $\kappa$ which we assume constant and their associated input fields $a_\text{in}$ and $b_\text{in}$, respectively.
With an input field in vacuum, the equations for the corresponding field amplitudes become
\begin{equation}
    \dod{\alpha}{t} = - \frac{(\gamma + \kappa)}{2}\alpha.
\end{equation}
This can be solved easily, yielding
\begin{equation}
    \alpha(t) = \alpha(0) \expup{-\frac{\kappa}{2} t  - \frac{1}{2}\int_0^t \gamma(t')\dif t'}.
\end{equation}
As shown earlier, after each round we have
\begin{equation}
    \begin{aligned}
        \langle \sigma_-(t)\rangle &= -i\sqrt{\gamma \Delta t}\alpha(t)\\
        &= -i\sqrt{\gamma(t) \Delta t} \expup{-\frac{\kappa}{2} t  - \frac{1}{2}\int_0^t \gamma(t')\dif t'} \times \alpha(0)
    \end{aligned}
\end{equation}
In the following, we consider without loss of generality that we perform a homodyne measurement of the $p$ quadrature coordinate, corresponding to a measurement of $\sigma_x$ at each round.
Assuming a qubit measurement fidelity of $\eta_m$, at each round we have $E[X_n] = \eta_m \langle \sigma_x(t)\rangle$, we thus have
\begin{equation}
    J = \sum_{n=1}^{\nmeas} f(t_n) X_n.
\end{equation}
To find the best mode-matching condition, we want to find the function $f(t)$ that minimizes the variance of $J$ under the constraint that $E[J] =  p(0)$. This constraint yields
\begin{equation}
    \begin{aligned}
    E[J] &= \sum_{n=1}^{\nmeas} f(t_n) \eta_m \langle \sigma_x(t)\rangle,\\
    &=  \eta_m\sum_n f(t_n) \sqrt{2\gamma(t_n) \Delta t} \expup{-\frac{\kappa}{2} t  - \frac{1}{2}\int_0^t \gamma(t')\dif t'} \times p(0).
    \end{aligned}
\end{equation}
Going to the continuous limit and defining $f(t_n) = \sqrt{\gamma(t_n) \Delta t }h(t_n)$, we obtain an equation independent of $p(0)$ for $h(t)$:
\begin{equation}\label{eq:ConstraintexpJ}
    \begin{aligned}
    1 &= \eta_m \int_0^\infty \dif t h(t) \sqrt{2}\gamma(t) \expup{-\frac{\kappa}{2} t  - \frac{1}{2}\int_0^t \gamma(t')\dif t'},\\
    \end{aligned}
\end{equation}
The variance of $J$ is given by
\begin{equation}
    \begin{aligned}
    E[J^2] - E[J]^2 &= \sum_n f(t_n)^2,\\
    &\rightarrow \int_0^\infty \dif t \gamma(t) h(t)^2.
    \end{aligned}
\end{equation}
Using variational calculus with Lagrange multipliers to take into account the constraint equation \eqref{eq:ConstraintexpJ}, we obtain
\begin{equation}
    \begin{aligned}
     h(t) \propto \expup{-\frac{\kappa}{2} t  - \frac{1}{2}\int_0^t \gamma(t')\dif t'},
    \end{aligned}
\end{equation}
which corresponds to
\begin{equation}
    \begin{aligned}
     f(t_n) = \frac{1}{\eta_m}\sqrt{\frac{\gamma(t_n) \Delta t}{2}}\frac{\expup{-\frac{\kappa}{2} t  - \frac{1}{2}\int_0^t \gamma(t')\dif t'}}{\int_0^\infty \dif \tau\gamma(\tau) \expup{-\kappa \tau  - \int_0^\tau \gamma(t')\dif t'}}.
    \end{aligned}
\end{equation}
Using Chebychev's inequality, we compute that
\begin{equation}
    \begin{aligned}
     \mathrm{Prob}[|J - p(0)|\geq \delta] \leq \frac{E[J^2] - E[J]^2}{\delta^2}.
    \end{aligned}
\end{equation}
We note that the above bound does not improve with the number of measurement $\nmeas$. Indeed, the qubitdyne measurement is a destructive measurement; once the cavity state is empty there is no more information to gain. As a result, one should in practice choose the smallest number $\nmeas$ where the conditions laid out in the main text are respected. Moreover, we remark that the bound above cannot be improved arbitrarily. This is because the measurement of the qubits correspond to some phase-preserving amplification, which limits the noise to a minimum of $1/2$.

In the constant $\gamma(t) = \gamma$ case, the filter function reduces to
\begin{equation}
    \begin{aligned}
     f(t_n) = \frac{1}{\eta_m}\sqrt{\frac{\gamma \Delta t}{2}}\frac{\kappa + \gamma}{\gamma} \expup{-\frac{(\kappa + \gamma)}{2} t},
    \end{aligned}
\end{equation}
which corresponds to a variance of
\begin{equation}
    \begin{aligned}
E[J^2] - E[J]^2 &= \frac{1+\kappa/\gamma}{2\eta_m^2}
    \end{aligned}
\end{equation}
which in turns bounds the error on the measurement
\begin{equation}
    \begin{aligned}
     \mathrm{Prob}[|J - p(0)|\geq \delta] \leq \frac{1+\kappa/\gamma}{2\eta_m^2 \delta}.
    \end{aligned}
\end{equation}

\section{Enhanced readout speed}
\begin{figure}[h!]
    \centering
    \includegraphics[width=\columnwidth]{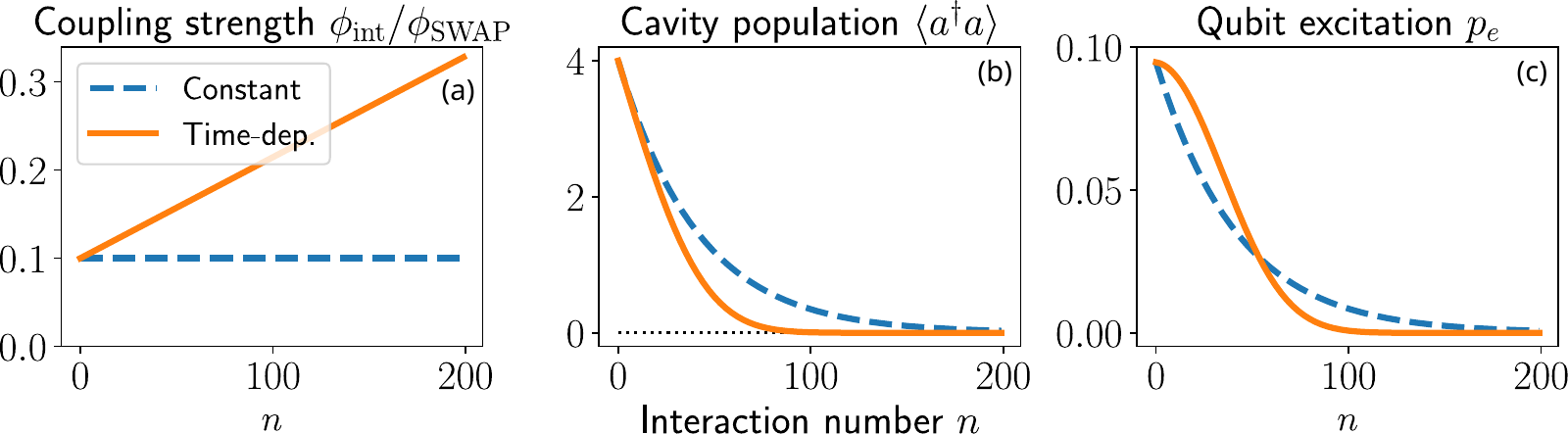}
    \caption{
    The effect of a time-dependent coupling on system populations compared to constant coupling for each step $n$ of a measurement round with $\nmeas=200$. (a) Constant interaction strength $\angle=0.1\swapangle$ (solid lines) and linearly increasing interaction $\angle(n)=0.1\swapangle + 0.0018n$ (dashed lines). (b) Cavity population with initial cat state $\alpha=2$. The dotted line indicates vacuum. Less than half the number of measurements are needed to empty the cavity for the time-dependent coupling. (c) The maximum qubit excitation probability is below 0.1.
    }
    \label{fig:compare_constant_timedep}
\end{figure}

Figure~\ref{fig:compare_constant_timedep} illustrates the reduction in measurements a time dependent coupling can provide, showing that the cavity with a $\alpha=2$ cat state can be emptied in less than 100 measurements for a linearly increasing coupling, while a constant $\angle=0.1\swapangle$ takes nearly 200 measurements. An increasing coupling strength thus reduces the required measurement time. Measurement statistics are accurate in both cases; state reconstruction with 10 homodyne angles gives 0.99 fidelity in both cases.
\section{Impact of qubit readout error}

\begin{figure}[h!]
    \centering
    \includegraphics[width=0.5\columnwidth]{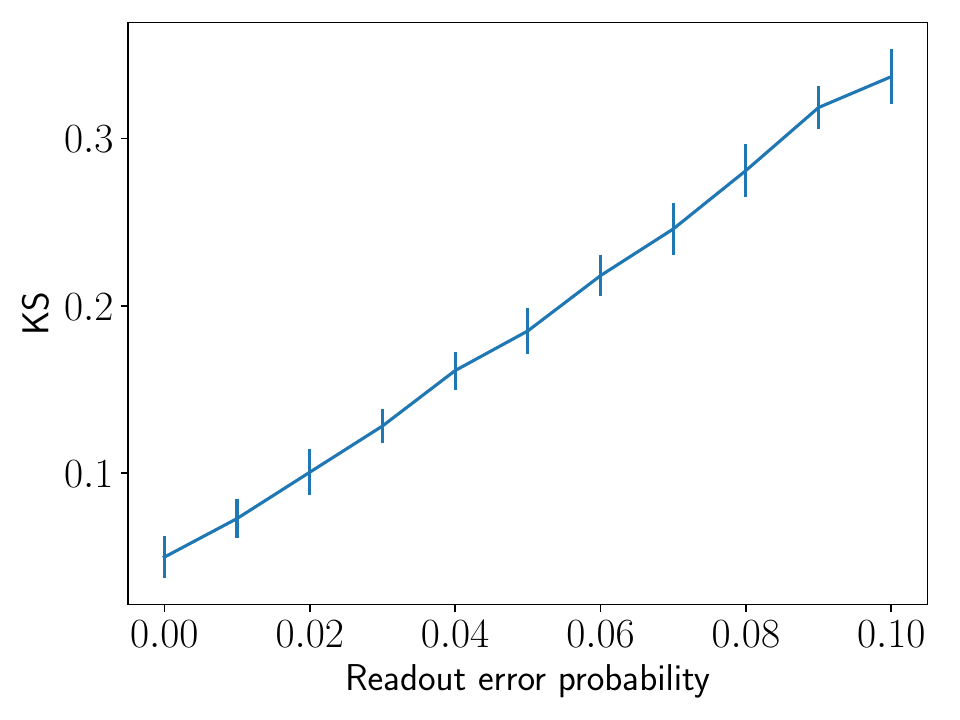}
    \caption{Kolmogorov-Smirnov statistic (KS) for different qubit readout error probabilities. Error bars indicate standard deviations for 20 runs.}
    \label{fig:ks_vs_pqerr}
\end{figure}

Qubit readout errors degrade the obtained measurement statistics. Fig.~\ref{fig:ks_vs_pqerr} displays the Kolmogorov-Smirnov statistic as a function of qubit readout error probability, again for the example cat state $\alpha=2$, using $\nmeas=200$ and $\ntrajs=1000$ in Fig.~\ref{fig:ks_vs_pqerr}. Each data point is an average over 20 full measurement records.

High qubit readout fidelity up to \SI{99.5}{\%} is attainable even without a quantum-limited amplifier~\cite{Chen2023Mar}, and this low error has very little effect on the average statistics. Hence, we neglect this type of error in the data analysis.

The qubit readout fidelity is equal to the probability to correctly measure the qubit state. In the qubitdyne protocol, qubit readout errors correspond to vacuum admixture in the detected state, which in optical detection typically arises from nonideal mode-matching or photodetector efficiency. This corresponds to the measured distribution being convolved with a Gaussian distribution centered at zero. For a coherent state $\ket{\alpha}$, this results in the distribution remaining Gaussian with the same variance, but moving towards the origin as $\alpha \rightarrow \sqrt{\eta}\alpha_0$ for detection efficiency $\eta$. This can be used to calibrate the efficiency~\cite{Pfaff2017Sep}. By simulating the center of the distribution of a coherent state for different qubit readout fidelities we can extract the corresponding qubitdyne detection fidelity $\eta_q$. For a fidelity \SI{99.5}{\%} the efficiency is $\eta_q=0.98$.
The code can be found in Ref.~\cite{code}.

\begin{figure}[h!]
    \centering
    \includegraphics[width=0.5\columnwidth]{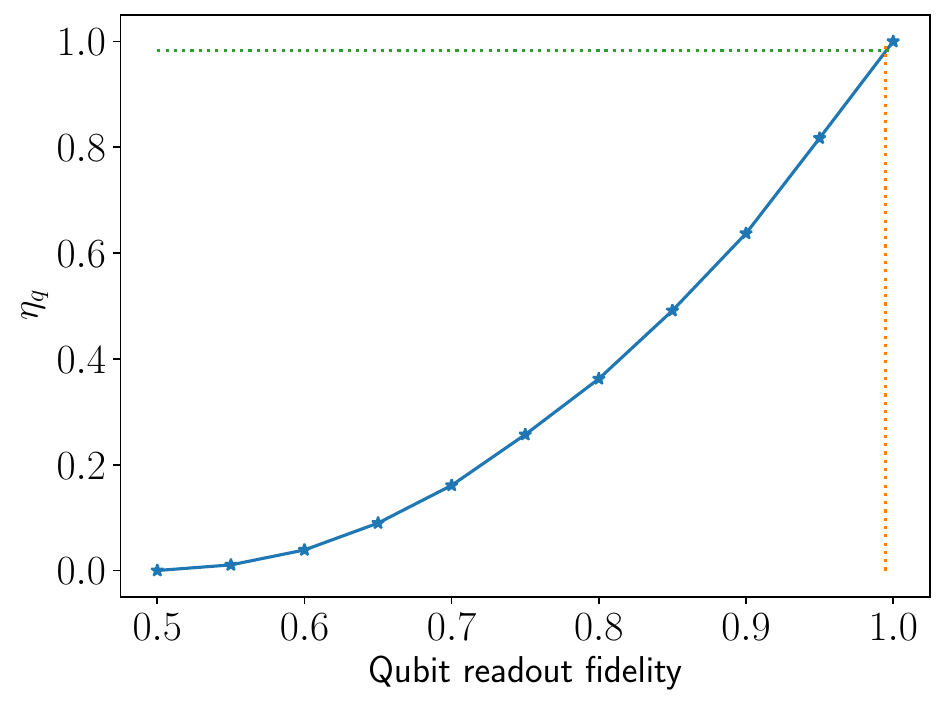}
    \caption{Qubitdyne efficiency $\eta_q$ resulting from imperfect qubit readout. The vertical dotted line indicates fidelity \SI{99.5}{\%}, and the horizontal dotted line indicates the corresponding efficiency $\eta_q=0.98$ as a visual aid.}
    \label{fig:enter-label}
\end{figure}
\section{Simulation parameter selection}

\subsection{Final cavity population}
\begin{figure}[h!]
    \centering
    \includegraphics[width=\columnwidth]{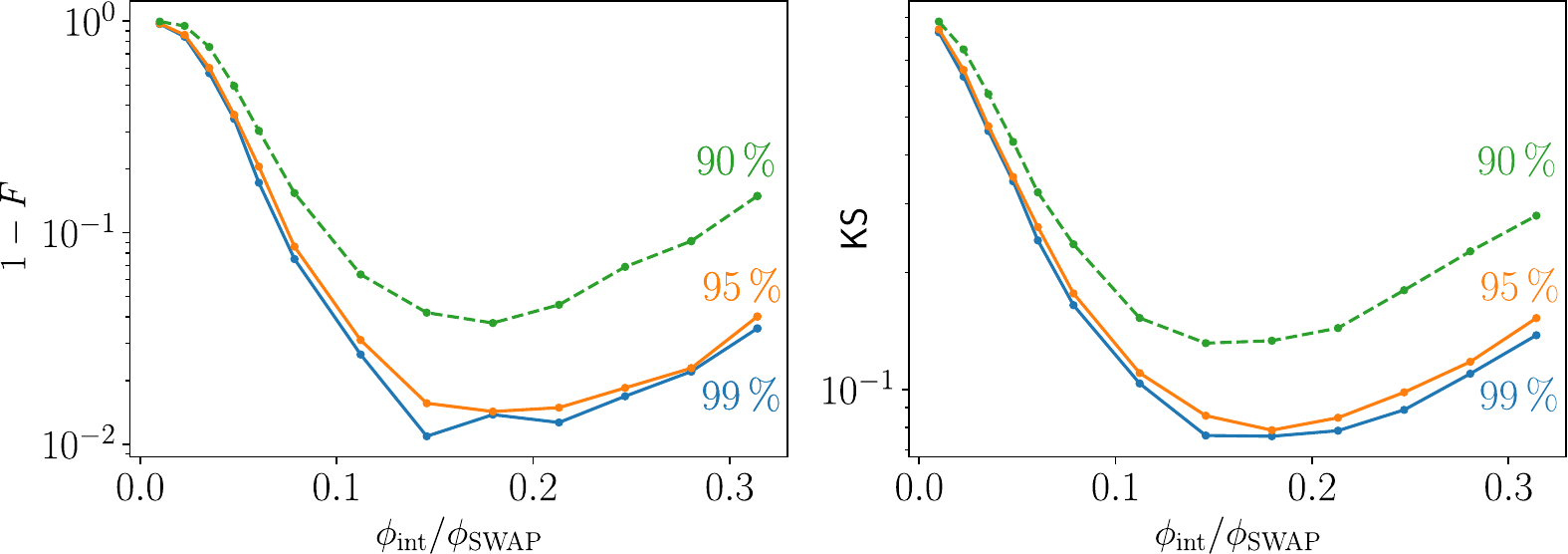}
    \caption{Three different realizations of infidelity and Kolmogorov-Smirnov statistic (KS) for different interaction strengths, using a coherent state $\alpha=4$. The percentage of final vacuum population is indicated, and it can be seen that \SI{95}{\%} vacuum is only marginally worse than \SI{99}{\%}, while \SI{90}{\%} is notably deviating. The simulation used $\ntrajs=1000$ trajectories with $\nmeas=200$ qubit measurements.}
    \label{fig:check_empty_pop_fid}
\end{figure}

As an example, for the coherent state $\alpha=4$, obtained infidelity $1-F$ and Kolmogorov-Smirnov (KS) statistics as a function of interaction strength are very similar when the cavity is emptied to \SI{99}{\%} and \SI{95}{\%} vacuum, but notably higher for \SI{90}{\%} vacuum. This is displayed in Fig.~\ref{fig:check_empty_pop_fid}. Thus, \SI{95}{\%} vacuum was chosen for Fig.\ 5 in the main text, as this provides reliable statistics with fewer simulated measurements than if the cavity was further emptied. This reduces simulation time, but would not speed up experiments unless the cavity can be reset faster than the qubitdyne protocol can empty it.

\subsection{Number of homodyne trajectories}

\begin{figure}[h!]
    \centering
    \includegraphics[width=0.5\columnwidth]{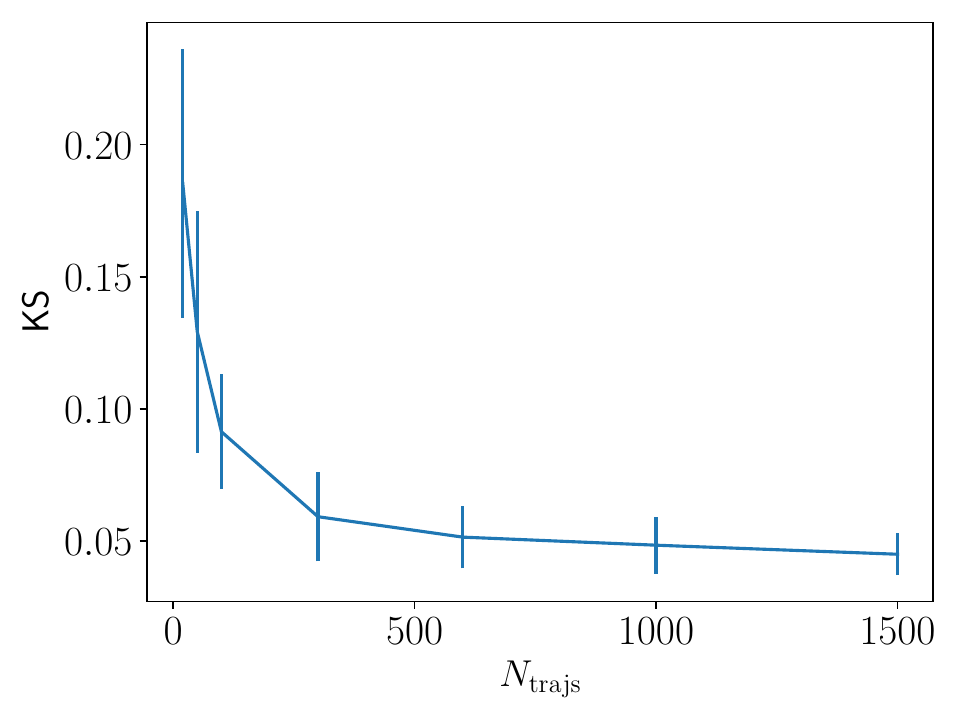}
    \caption{Kolmogorov-Smirnov statistic (KS) for different numbers of quantum trajectories $\ntrajs$ for a coherent state $\alpha=4$. Error bars indicate standard deviations for 20 runs. The KS value has largely converged after $\ntrajs=1000$.}
    \label{fig:ks_vs_ntrajs}
\end{figure}

The number of trajectories $\ntrajs$ was chosen such that using more would not result in better statistics on average. Fig.~\ref{fig:ks_vs_ntrajs} shows the Kolmogorov-Smirnov statistic for measurements of one quadrature of the cat state $\alpha=2$, using $\nmeas=200$ qubit measurements, for different numbers of realized trajectories. Due to fluctuations in the random measurement outcomes, the plot shows the average of 20 instances of each $\ntrajs$ value.

\section{Experimental implementations with superconducting circuits}

The most straightforward way to perform the required controllable SWAP interaction with a superconducting cavity is to repeatedly put it in resonance with a flux-tunable qubit~\cite{Hofheinz2009May}. However, if frequency tunability is challenging, such as when coupling to 3D cavities, a dispersively coupled fixed-frequency qubit can be used by driving the partial SWAP operation through various pumping schemes~\cite{Zeytinoglu2015Apr, Campagne-Ibarcq2018May, Kudra2022Dec,zhou2023realizing}. 
Specifically, one alternative to generate a tunable coupling between a cavity mode and a fixed frequency superconducting qubit in the strong dispersive regime is to apply a microwave drive at the energy difference between the states $\ket{f0}$ and $\ket{g1}$. At higher photon numbers, the resonant frequencies of transitions $\ket{f,n}\leftrightarrow\ket{g,n+1}$ are shifted due to the photon-number-dependent dispersive shifts \cite{gasparinetti2016measurement}. By driving these transitions with a comb of drives, superposed oscillating signals are produced, corresponding to each populated Fock level in the cavity. This results in an effective resonant swap between the qubit and the cavity.

\subsection{SWAP interaction through controlled displacements}
Alternatively, a SWAP-type interaction can also be well approximated using controlled-displacements~\cite{Eickbusch2022Dec}.
In general, any qubit-cavity unitary can be realized by a decomposition into qubit rotations and conditional displacements of the cavity~\cite{Eickbusch2022Dec}. Here, we can analytically compute a simple decomposition of this form for Eq.\ (1) from the second-order Suzuki-Trotter formula
\begin{equation}
    \begin{aligned}
        U &= \exp\{-i\angle\sqrt 2 (x \sigma_x + p \sigma_y) \},\\
        &\approx e^{-i\frac{\angle}{\sqrt 2} x \sigma_x}e^{-i \angle\sqrt 2 p \sigma_y }e^{-i\frac{\angle}{\sqrt 2} x \sigma_x } + \mathcal O(\angle^3).
    \end{aligned}
\end{equation}
This equation corresponds to a series of cavity displacements conditioned on the qubit states in the $\sigma_x,\sigma_y$ bases, and can easily be converted into the framework of Ref.~\cite{Eickbusch2022Dec}. We note that repeating such an interaction was used in Ref.~\cite{Sivak2023Apr} to reset the cavity in its ground state, although the qubit measurement results were not analyzed there.

\section{Homodyne measurement through phase estimation}
As mentioned in the main text, homodyne measurements can also be realized through phase estimation of $e^{i \epsilon x_\theta}$. The phase is $\hat \phi = \epsilon \hat x_\phi$, and
 $\tilde \phi$ is its estimated value, with $\tilde x_\phi = \tilde \phi /\epsilon$ being the estimated value of $x_\phi$.
Choosing $\epsilon$ small enough such that the initial state to be measured only has support within $x_\theta \in [-\pi/\epsilon,\pi/\epsilon]$, a measurement of $\hat x_\phi\; \mathrm{mod}\; 2\pi/\epsilon$ is equivalent to a measurement of $\hat x_\phi$.
Code for numerical simulations can be found in Ref.~\cite{code}.
Phase estimation is a problem with a lot of history, below we follow Ref.~\cite{Tehral2016Jan} to explain the procedure.

\subsection{Standard phase estimation}
\begin{figure}[h!]
    \centering
    \includegraphics[width=0.5\columnwidth]{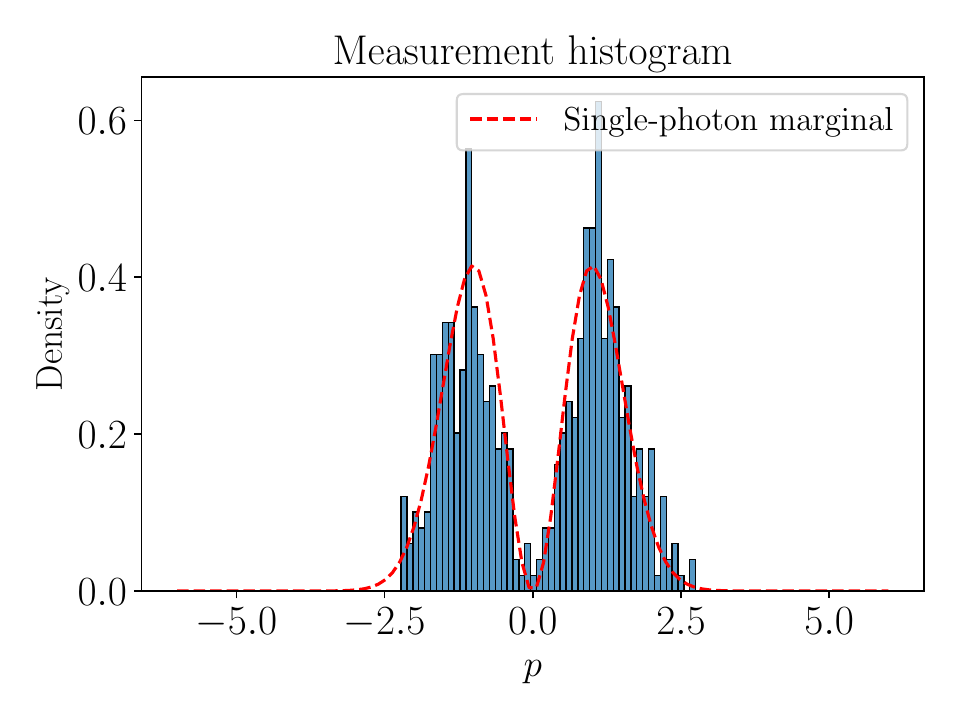}
    \caption{Simulated measurement histogram from qubit measurements after controlled displacement and qubit rotation, with initial cavity Fock state $\ket{1}$. The simulation used $N_m=100$ qubit measurements and $\ntrajs=500$ trajectories.}
    \label{fig:phase_est}
\end{figure}

Following the iterative quantum phase estimation protocol~\cite{nielsen_quantum_2010}, we can get an estimate for the phase, for which the probability that we make an error larger than $\delta$ is bounded by
\begin{equation}
    \mathrm{Prob}[|\tilde x_\phi - x_\phi|\geq \delta] \leq \frac{1}{2\epsilon 2^{N_m}\delta - 2},
\end{equation}
for $N_m$ measurements and an error $\epsilon 2^{N_m}\delta \geq 1$.
Considering that the maximum required displacement is given by $\epsilon_{\mathrm{max}} = \epsilon 2^{N_m-1}$, the probability of success only depends on the maximum displacement. However, note that with this method, the range of probed values is set by (the inverse of) the minimum displacement value.
A simulation  example is shown in Fig.~\ref{fig:phase_est}


\subsection{Non-adaptative phase estimation}
The simplest approach is to repeat the phase estimation circuit $N_m$ times, where the real and imaginary parts of $e^{i \epsilon x_\theta}$ are measured $N_m$ times each. Assuming that the state is prepared in an eigenstate of $x_\theta$, we obtain that at each round the probability of measuring the qubit in the ground state is
\begin{equation}
\begin{aligned}
    p_\mathrm{R} &\equiv P_\mathrm{Real}(g|x_\theta) = \frac{1 + \cos(\epsilon x_\theta)}{2},\\
    p_\mathrm{I} &\equiv P_\mathrm{Imag}(g|x_\theta) = \frac{1 + \sin(\epsilon x_\theta)}{2}.
    \end{aligned}
\end{equation}
We can estimate the angle from the fraction $g$ of results for the real part $\tilde p_R$ and imaginary part $\tilde p_I$, from which we obtain
\begin{equation}
    \tilde \phi = \mathrm{arg}[2\tilde p_R - 1 +i(2\tilde p_I -1)],
\end{equation}
with an estimate for the measurement value $\tilde x = \tilde \phi/\epsilon$. Through Chernoff's bound, we can find
\begin{equation}
    \mathrm{Prob}[|\tilde x_\theta - x_\theta|\geq \delta] \leq 4 e^{-\frac{\sin(\delta/\epsilon)N_m}{2\sqrt 2}}.
\end{equation}

\subsection{Adaptative phase estimation}
In the method above, we real and imaginary parts of $e^{i \epsilon x_\theta}$ are measured in turn, which in practice is obtained by adding a phase of $\varphi = 0$ or $\varphi = \pi/2$ to the qubit before applying a Hadamard gate and reading it out.
Alternatively, we can adapt the measurement phase at each step to optimize the amount of information learned. Starting with a phase of $\varphi_1 = 0$, we can choose the measurement phase at the $m$th round from
\begin{equation}
    \varphi_m = \mathrm{arg max}\sum_{x_m = 0,1}\left|\int d\phi e^{i\phi}P_\varphi(x[m]|\phi) \right|,
\end{equation}
with
\begin{equation}
    P_\varphi(x[m]|\phi) = \prod_{i=1}^m \cos^2\left(\frac{\phi + \varphi_i + x_i \pi}{2}\right).
\end{equation}

We choose the phase estimate from
\begin{equation}
    \tilde \phi = \mathrm{arg}\int d\phi\, e^{i\phi}P_\varphi(x[m]|\phi).
\end{equation}

\renewcommand{\theequation}{S\arabic{equation}}
\renewcommand{\thefigure}{S\arabic{figure}}
\renewcommand{\bibnumfmt}[1]{[S#1]}
\renewcommand{\citenumfont}[1]{S#1}

\newpage
\bibliographystyle{apsrev4-1}